\documentclass[aps,pra,showpacs,twocolumn,a4paper]{revtex4}

\usepackage{amsmath,amsfonts,amssymb}
\usepackage{txfonts}
\usepackage{type1cm}

\begin{document}

\bibliographystyle{apsrev}

\title{Additivity and multiplicativity properties of some Gaussian channels for Gaussian inputs}
\author{Tohya Hiroshima}
\email{tohya@qci.jst.go.jp}
\affiliation{
Quantum Computation and Information Project, ERATO-SORST, Japan Science and Technology Agency,\\
Daini Hongo White Building 201, Hongo 5-28-3, Bunkyo-ku, Tokyo 113-0033, Japan
}
\date{\today}
%%%%%%%%%%%%%%%%%%%%%%%%%%%%%%%%%%%%%%%%%%%%%%%%%%%%%%%%%%%%%%%%%%%%%%%%%%%%%%%
\begin{abstract}
We prove multiplicativity of maximal output $p$ norm of classical noise channels and 
thermal noise channels of arbitrary modes for all $p>1$ 
under the assumption that the input signal states are Gaussian states.
As a direct consequence, we also show the additivity of the minimal output entropy 
and that of the energy-constrained Holevo capacity for those Gaussian channels 
under Gaussian inputs.
To the best of our knowledge, newly discovered majorization relation on symplectic eigenvalues, 
which is also of independent interest, plays a central role in the proof.
\end{abstract}
%%%%%%%%%%%%%%%%%%%%%%%%%%%%%%%%%%%%%%%%%%%%%%%%%%%%%%%%%%%%%%%%%%%%%%%%%%%%%%%
\maketitle
%%%%%%%%%%%%%%%%%%%%%%%%%%%%%%%%%%%%%%%%%%%%%%%%%%%%%%%%%%%%%%%%%%%%%%%%%%%%%%%

\section{Introduction}

One of the goals of quantum information theory is to clarify the ultimate capability of information processing harnessed by using quantum mechanics \cite{NC,Hay}.
The celebrated Holevo-Schumacher-Westmoreland theorem \cite{Hol98a,SW} gives us 
a formal basis to determine the ultimate transmission rate of classical information 
encoded in quantum states transmitted through a quantum channel.
Yet, an important question is still unanswered 
in terms of the classical capacity of quantum channels.
It is the additivity question; 
Do the entangled inputs over several invocations of quantum channels improve 
the classical capacity of quantum channels?
Despite many efforts devoted to the additivity problems of quantum channels, 
the additivity properties have been proven for a few examples, such as 
entanglement breaking channels \cite{Sho02}, unital qubit channels \cite{Kin02}, 
depolarizing channels \cite{Kin03}, and contravariant channels \cite{MY}.
Surprisingly, the additivity problems of quantum channels have been shown to be equivalent to 
the seemingly unrelated additivity problems of quantum entanglement, 
i.e., the additivity and the strong superadditivity 
of entanglement of formation \cite{MSW,AB,Sho04,Pom}.
All of them are not completely solved and are now major concerns 
in quantum information and the quantum entanglement theories.

As for continuous-variable quantum systems, in spite of intensive research \cite{GL,GLM,GGL04a}, 
only lossy channels have been proven to be additive \cite{GGL04b}.
The additivity problems may be much more intractable for continuous-variable quantum systems.
The natural question is, therefore, what can we say about the additivity properties 
of Gaussian quantum channels if we restrict the input signal states to be Gaussian states?
This question has its own significance.
One rationale is that the Gaussian channels correspond to the so-called Gaussian operations 
that can be implemented by current experimental techniques, 
such as bemsplitters, phase shifters, squeezers, and homodyne measurements.
Another is the mathematical simplicity; 
Gaussian operations on Gaussian states are completely characterized by 
finite dimensional matrices and vectors, 
although the underlying Hilbert space is infinite dimensional.
Due to their mathematical simplicity, 
the additivity problems of Gaussian channels under Gaussian inputs provide a potential firm step 
towards answering the additivity questions.

Serafini {\it et al.} \cite{SEW} formulated the multiplicativity problems 
of the purity at the output of Gaussian channels measured by the Schatten $p$ norm 
under the assumption that the input signal states were Gaussian states.
In this paper, we extend their formalism to the additivity problems of minimal output entropy 
and energy-constrained Holevo capacity 
and prove the additivity properties of two classes of Gaussian channels --- 
the classical noise channels and thermal noise channels of arbitrary modes.

The paper is organized as follows.
In Sec.~\ref{Gaussian_states} we introduce the notation and present basic facts 
about Gaussian states and the symplectic transformations used in this paper.
In Sec.~\ref{Gaussian_channels} we define Gaussian channels and 
introduce three figures of merits to quantify Gaussian channels --- 
the maximal output $p$ norm, the minimal output entropy, and the Holevo capacity.
In Sec.~\ref{additivity_problems} we formulate the additivity and multiplicativity problems 
of Gaussian channels for Gaussian inputs.
In Sec.~\ref{majorization} we prove a trace formula for symplectic eigenvalues 
and a majorization relation on symplectic eigenvalues 
that is an immediate consequence of the trace formula.
By virtue of this majorization relation on symplectic eigenvalues, 
we prove the additivity and multiplicativity properties of classical noise channels 
and thermal noise channels of arbitrary modes in Sec.~\ref{additivity_proof}.
Section~\ref{conclusion} is devoted to concluding remarks.

\section{Gaussian states} \label{Gaussian_states}

In this section, we introduce the notation and summarize 
the basic facts about Gaussian states and symplectic transformations \cite{EP}.
We consider an $n$ mode quantum system, such as a radiation field.
Each mode corresponds to a quantum mechanical harmonic oscillator 
with two canonical degrees of freedom and the quadratures of each mode correspond to the position and momentum of the harmonic oscillator.
Thus an $n$ mode state has a $2n$ canonical degrees of freedom.
Let $Q_{k}$ and $P_{k}$ denote the ``position'' and ``momentum'' operators associated with 
the $k$th mode $(k=1,2,\ldots ,n)$.
These operators or canonical variables are written 
in terms of the creation and annihilation operators of the mode;
\begin{equation}
Q_{k}=\sqrt{\frac{1}{2\omega _{k}}}\left( a_{k}+a_{k}^{\dagger }\right) 
\end{equation}
and
\begin{equation}
P_{k}=-i\sqrt{\frac{\omega _{k}}{2}}\left( a_{k}-a_{k}^{\dagger }\right), 
\end{equation}
where $\omega _{k}$ denotes the energy of the $k$th mode ($\hbar =1$).
Since $[a_{j},a_{k}]=[a_{j}^{\dagger },a_{k}^{\dagger }]=0$ and 
$[a_{j},a_{k}^{\dagger }]=\delta _{jk}$, 
we have $[Q_{j},Q_{k}]=[P_{j},P_{k}]=0$ and $[Q_{j},P_{k}]=i\delta _{jk}$.
Defining
\begin{eqnarray}
R &=&(R_{1,}R_{2},\ldots ,R_{2n})^{T} \nonumber \\
&=&(\omega _{1}^{1/2}Q_{1},\omega _{1}^{-1/2}P_{1},\ldots ,\omega
_{n}^{1/2}Q_{n},\omega _{n}^{-1/2}P_{n})^{T},
\end{eqnarray}
these canonical commutation relations (CCRs) can be written as 
$[R_{j},R_{k}]=i(J_{n})_{jk}$.
Here, $J_{n}=\oplus _{j=1}^{n}J_{1}$ with
\begin{equation}
J_{1}=\left( 
\begin{array}{cc}
0 & 1 \\ 
-1 & 0
\end{array}
\right).
\end{equation}
In the following, the characteristic function defined as 
$\chi (\xi )={\rm Tr}[\rho \mathcal{W}(\xi )] $ 
plays a key role.
Here, $\mathcal{W}(\xi )=\exp (i\xi ^{T}J_{n}R)$ is called the Weyl operators and $\rho $ denotes the density opeartor.
The density operator in turn can be written in terms of its characteristic function 
and Weyl operators as follows.
\begin{equation}
\rho =\frac{1}{(2\pi )^{n}}\int d^{2n} \xi \chi (\xi )\mathcal{W}(-\xi ).
\end{equation}

A Gaussian state is defined as a state whose characteristic function is a Gaussian function:
\begin{equation} \label{eq:characteristic}
\chi (\xi )=\exp [ -\frac{1}{4}\xi ^{T}\Gamma \xi +iD^{T}\xi ].
\end{equation}
Here, $\Gamma > 0$ is a real symmetric matrix and $D\in \mathbb{R}^{2n}$.
The first moment is also called the displacement or mean and given by 
$m_{j}={\rm Tr}(\rho R_{j})$ and the second moment is given by
\begin{equation} \label{eq:covariance}
\gamma _{jk}=2{\rm Tr}[\rho (R_{j}-m_{j})(R_{k}-m_{k})]-i(J_{n})_{jk},
\end{equation}
which is called the covariance of canonical variables.
The $2n \times 2n$ real symmetric matrix $(\gamma _{jk}) $ is called 
the covariance matrix $\gamma $.
$\Gamma $ and $D$ in Eq.~(\ref{eq:characteristic}) are given by 
$\Gamma =J_{n}^{T} \gamma J_{n}$ and $D=J_{n} m$.

Note that due to our choice of canonical varibles, 
$R_{2j-1}=\omega _{j}^{1/2}Q_{j}$ 
and 
$R_{2j}=\omega _{j}^{-1/2}P_{j}$ 
($j=1,2,\ldots ,n$), 
the trace of the principal submatrix of the $j$th mode of $\gamma $,
\begin{equation} \label{eq:submatrix}
\gamma _{[j]}=\left( 
\begin{array}{cc}
\gamma _{2j-1,2j-1} & \gamma _{2j-1,2j} \\ 
\gamma _{2j,2j-1} & \gamma _{2j,2j}
\end{array}
\right)
\end{equation}
gives the energy of the $j$th mode if $m_{2j-1}=m_{2j}=0$;
\begin{equation}
\frac{1}{4}\omega _{k}\left( \gamma _{2j-1,2j-1}+\gamma _{2j,2j}\right)
=\omega _{k}\left( \left\langle a_{j}^{\dagger }a_{j}\right\rangle +\frac{1}{%
2}\right). 
\end{equation}

A density operator $\rho $ is a positive semidefinite operator ($\rho \geq 0$) with 
${\rm Tr}\rho =1$.
The necessary and sufficient condition for $\rho \geq 0$ in a Gaussian state is given 
in terms of the covariance matrix as follows \cite{Pet}.
\begin{equation} \label{eq:physical}
\gamma +iJ_{n} \geq 0.
\end{equation}
Furthermore, the necessary and sufficient condition for a pure Gaussian state is given by \cite{MV}
\begin{equation} \label{eq:pure}
\det \gamma =1.
\end{equation}

A linear transformation on canonical variables is written as $R\rightarrow R^{\prime }=SR$.
Since the new variables $R ^{\prime } $ also must conserve the CCR 
$[R_{j}^{\prime },R_{k}^{\prime }]=i(J_{n})_{jk}$, $SJ_{n}S^{T}=J_{n}$ must hold.
Such an $2n \times 2n$ real matrix satisfying $SJ_{n}S^{T}=J_{n}$ is called 
a symplectic transformation, $S\in {\rm Sp}(2n,\Bbb{R})=\{S|SJ_{n}S^{T}=J_{n}\}$, 
which forms a group so that $S^{-1}$ and $S_{1}S_{2}$ are symplectic 
if $S,S_{1},S_{2}\in {\rm Sp}(2n,\Bbb{R})$.
Furthermore, $S^{T}$ is also symplectic and $\det S=1$ \cite{SSM} if $S \in {\rm Sp}(2n,\Bbb{R})$.
In this paper, we repeatedly use the following Williamson theorem \cite{Wil}.
For a real symmetric positive definite $2n\times 2n$ matrix $A=A^{T}>0$, 
there exists a symplectic transformation $S\in {\rm Sp}(2n,\mathbb{R})$ such that
\begin{equation} \label{eq:Williamson}
SAS^{T}=\mathrm{diag}(\nu _{1},\nu _{1},\nu _{2},\nu _{2},\ldots,\nu _{n}\nu _{n})
\end{equation}
with $\nu _{j}(>0)$ called symplectic eigenvalues of $A$ $(j=1,2,\ldots ,n)$.
Equation~(\ref{eq:Williamson}) is called the Williamson standard form of $A$.
The symplectic eigenvalues can be computed via the eigenvalues of $J_{n}A$, which are 
$\pm i\nu _{j}$ ($j=1,2,\ldots ,n$).
Any symplectic transformation $S\in {\rm Sp}(2n,\mathbb{R})$ can be decomposed into
\begin{equation} \label{eq:Euler}
S=T^{(1)}ZT^{(2)}
\end{equation}
with $T^{(1)},T^{(2)}\in {\rm Sp}(2n,\mathbb{R})\cap O(2n) = K(n) $ and
\begin{equation}
Z=\mathrm{diag}(z_{1},z_{1}^{-1},\ldots ,z_{n},z_{n}^{-1}),
\end{equation}
where $z_{j}\geq 1$ ($j=1,2,\ldots,n$) \cite{Pramana}.
$O(2n)$ denotes the orthogonal group whose elements are $2n \times 2n$ real orthogonal matrices.
Equation~(\ref{eq:Euler}) is called the Euler decomposition of symplectic transformations.
$K(n) $ is a maximal compact subgroup of ${\rm Sp}(2n,\Bbb{R}) $ and is isomorphic to $U(n)$, 
the unitary group whose elements are $n \times n$ unitary matrices \cite{SMD}.
The isomorphism is established via the following correspondance:
\begin{equation} \label{eq:isomorphism_1}
T_{2j-1,2k-1}=T_{2j,2k}=\mathrm{Re}u_{j,k}=u_{j,k}^{R}
\end{equation}
and
\begin{equation} \label{eq:isomorphism_2}
T_{2j-1,2k}=-T_{2j,2k-1}=\mathrm{Im}u_{j,k}=u_{j,k}^{I},
\end{equation}
where $T \in K(n)$ and $u_{j,k}$ are $(j,k)$ components of $n \times n$ unitary matrices $U $.
Using Eqs.~(\ref{eq:isomorphism_1}) and (\ref{eq:isomorphism_2}), 
the isomorphism $K(n) \simeq U(n)$ is easily verified by direct calculations.

Since the covariance matrix of an $n$-mode state 
is a $2n\times 2n$ real symmetric positive-definite matrix, 
it can be cast into the Williamson standard form.
In terms of symplectic eigenvalues, 
condition (\ref{eq:physical}) is rephrased as $\nu _{j}\geq 1$ ($j=1,2,\ldots ,n$) and 
condition (\ref{eq:pure}) is written as $\nu _{j}=1$ for all $j$.

A canonical linear transformation corresponds to a unitary transformation in the Hilbert space.
Such a unitary transformation is defined by 
$U_{S}\mathcal{W}(\xi )U_{S}^{\dagger }=\mathcal{W}(S^{-1}\xi )$, 
and the density operator is transformed as 
$\rho \mapsto U_{S}\rho U_{S}^{\dagger }=\widetilde{\rho }$ correspondingly.
It is easy to see that $U_{S}^{\dagger }=U_{S^{-1}}$.
The characteristic function of the new state $\widetilde{\rho }$ is given by 
${\rm Tr}[U_{S}\rho U_{S}^{\dagger }\mathcal{W}(\xi )] =
{\rm Tr}[\rho \mathcal{W}(S \xi )]=\chi (S \xi )$.
Accordingly, the covariance matrix and the displacement are transformed as 
$\gamma \mapsto S^{-1}\gamma (S^{-1})^{T}$ and $m\mapsto S^{-1}m$.
Note that the symplectic eigenvalues are invariant 
under such symplectic transformations on the covariance matrix.

Coherent states, squeezed states, and thermal states are typical Gaussian states, 
while the number states (of the single mode) given by 
$\left| k\right\rangle \left\langle k\right| $ 
with
\begin{equation}
\left| k\right\rangle =\frac{1}{\sqrt{k!}}(a^{\dagger })^{k}\left|
0\right\rangle, \qquad k=1,2,\ldots,
\end{equation}
are not.
However, a vacuum state 
$\left| 0\right\rangle \left\langle 0\right| $, 
which is a special case of the number states, is a Gaussian state with the covariance matrix,
\begin{equation} \label{eq:vacuum_covariance}
\gamma _{vac}=\mathrm{diag}(1,1).
\end{equation}
This is the minimal-energy pure state.
The coherent state is the displaced vacuum state 
so that the covariance matrix is given by Eq.~(\ref{eq:vacuum_covariance}) 
but has a finite displacement.
The thermal state of the single mode,
\begin{equation}
\rho _{th}=\frac{1}{1+\left\langle n\right\rangle }\sum_{k=0}^{\infty
}\left( \frac{\left\langle n\right\rangle }{1+\left\langle n\right\rangle }%
\right) ^{k}\left| k\right\rangle \left\langle k\right| 
\end{equation}
has the covariance matrix
\begin{equation} \label{eq:thermal_covariance}
\gamma _{th}=\mathrm{diag}(2\left\langle n\right\rangle +1,2\left\langle
n\right\rangle +1)
\end{equation}
with $\left\langle n\right\rangle $ being the mean photon number of the mode.

\section{Gaussian channels and their quantification} \label{Gaussian_channels}

A Gaussian channel $\Phi $ is a completely positive trace preserving map 
that maps Gaussian input states $\rho $ to Gaussian output states $\Phi (\rho) $ \cite{HW,EW}.
The covariance matrix is transformed according to
\begin{equation} \label{eq:channel_covariance}
\gamma \mapsto \phi (\gamma )=X^{T}\gamma X+Y,
\end{equation}
where $X$ and $Y$ are $2n \times 2n$ real matrices and 
$Y$ is positive and symmetric ($Y = Y^{T} \geq 0$).
The complete positivity of the channel is expressed in terms of these matrices as \cite{Lin}
\begin{equation}
Y+iJ_{n}-iX^{T}J_{n}X\geq 0.
\end{equation}

Hereafter, we write a Gaussain channel by a capital greek letter and 
the coresponding transformation on the covariance matrix 
by the corresponding lower case greek letter.

There are several figures of merits for quantifying quantum channels.
Here we take three of them; 
the maximal output $p$ norm \cite{AHW}, the minimal output entropy \cite{KR},
and the Holevo capacity \cite{Hol79} for Gaussian state inputs.

The Gaussian maximal output $p$ norm is defined as
\begin{equation} \label{eq:maximal_output_norm}
\xi _{p}(\Phi )=\sup_{\rho \in \mathcal{G}}\left\| \Phi (\rho )\right\| _{p}, 
\end{equation}
where 
$\left\| \rho \right\| _{p}=\left( {\rm Tr}\left| \rho \right| ^{p}\right)^{1/p} $ 
is the Schatten $p$ norm ($p\geq 1$) with $\left| A\right| =\sqrt{A^{\dagger }A}$.
In Eq.~(\ref{eq:maximal_output_norm}), $\mathcal{G}$ denotes the set of all Gaussian states.
For a Gaussian state with covariance matrix $\gamma $,
\begin{equation}
{\rm Tr}\rho _{\gamma}^{p}=\prod_{j=1}^{n}\frac{2^{p}}{f_{p}(\nu _{j})}=%
\frac{2^{pn}}{F_{p}(\nu )}, 
\end{equation}
where
\begin{equation}
f_{p}(x)=(x+1)^{p}-(x-1)^{p}. 
\end{equation}
This formula has been originally derived in \cite{HSH00}.
Note that ${\rm Tr}\rho _{\gamma}^{p}$ is independent of the displacement $m$.
We can verify that $\ln f_{p}$ is increasing and concave (Appendix A), 
so that $F_{p}(\nu )=\prod_{j=1}^{n}f_{p}(\nu _{j})$ 
is increasing and Schur-concave (Appendix B).
In Eq.~(\ref{eq:maximal_output_norm}), $\mathcal{G}$ can be replaced by 
$\mathcal{G}_{p}$, the set of all pure Gaussian states \cite{SEW}.
In terms of $F_{p}$, we have
\begin{equation}
\left( \frac{2^{n}}{\xi _{p}(\Phi )}\right) ^{p}=
\inf_{\gamma _{p}}F_{p}\left[\nu \left(\phi (\gamma _{p})\right)\right]. 
\end{equation}

The Gaussian minimal output entropy is defined as
\begin{equation} \label{eq:minimal_output_entropy}
S_{\min }(\Phi )=\inf_{\rho \in \mathcal{G}}S\left(\Phi (\rho )\right), 
\end{equation}
where $S(\rho )=-{\rm Tr}\rho \ln \rho $ is the von Neumann entropy.
Followed by the arguments presented in \cite{SEW}, it can be shown that 
$\mathcal{G}$ in Eq.~(\ref{eq:minimal_output_entropy}) can be replaced by $\mathcal{G}_{p}$.
Since
\begin{equation} \label{eq:derivative}
\lim_{p\rightarrow 1+}\frac{d}{dp}\left\| \rho \right\| _{p}=-S(\rho ),
\end{equation}
$S_{\min }(\Phi )$ can be computed through $\xi _{p}(\Phi )$.
Note that $S_{\min }(\Phi )$ is also independent of the displacement $m$.
Hereafter, we have occasions to write the von Neumann entropy as $S(\gamma )$ 
instead of $S(\rho _{\gamma })$ when we are dealing with Gaussian states.

By definition, the Holevo capacity for Gaussian state inputs or 
the Gaussian Holevo capacity is written as \cite{Hol98b}
\begin{equation} \label{eq:Gaussian_Holevo_capacity}
C_{G}(\Phi)=\sup_{\mu ,\rho _{(\gamma ,m)}}\left[ S\left(\Phi (\overline{\rho }%
)\right)-\int \mu (d\gamma ,dm)S\left(\Phi (\rho _{(\gamma ,m)})\right)\right], 
\end{equation}
where
\begin{equation}
\overline{\rho }=\int \mu (d\gamma ,dm)\rho _{(\gamma ,m)} 
\end{equation}
is the averaged signal state.
In Eq.~(\ref{eq:Gaussian_Holevo_capacity}), 
the supremum is taken over all possible probability measures $\mu $ 
and signal states $\rho _{(\gamma ,m)}$ constituting the signal ensemble.
Since the states $\rho _{(\gamma ,m)}$ are infinite dimensional states, 
the right-hand side of Eq.~(\ref{eq:Gaussian_Holevo_capacity}) becomes any large number 
if we do not impose some constraint on the signal states.
Here, we take the energy constraint,
\begin{equation} \label{eq:energy_constraint_1}
{\rm Tr}\overline{\rho }H=\mathcal{E}
\end{equation}
with
\begin{equation}
H =\sum_{k=1}^{n}\omega _{k}\left( a_{k}^{\dagger }a_{k}+\frac{1}{2}\right)
=\frac{1}{2}\sum_{k=1}^{n}\omega _{k}\left( R_{2k-1}^{2}+R_{2k}^{2}\right). 
\end{equation}
Here we recall that the von Neumann entropy of a Gaussian state depends only on the covariance matrix, 
and that channel $\Phi $ affects only the covariance matrix.
Therefore, if we find a single state $\rho _{(\gamma ^{*},m)}$ that minimizes $S\left(\Phi (\rho )\right)$,
all possible Gaussian states $\rho $ with the covariance matrix, $\gamma ^{*}$, also minimizes 
$S(\Phi (\rho ))$.
This observation indicates that the optimal signal ensemble 
that attains the Gaussian Holevo capacity consists of Gaussian states 
with the common covariance matrix $\gamma ^{*}$ and a certain probability distribution 
of the displacement $m$.
If we restrict the signal ensemble to that described above, 
it suffices to take a Gaussian probability distribution for the probality measure 
$\mu (d\gamma ,dm)=\mu (dm)$.
This is shown as follows \cite{HSH99}.
If $\mu (dm)$ is a Gaussian distribution;
\begin{equation}
\mu (dm)=\frac{1}{\pi ^{n}\sqrt{\det Y_{\mu }}}\exp (-m^{T}Y_{\mu }^{-1}m)dm
\end{equation}
with $Y_{\mu } > 0$, 
the averaged input signal state is calculated as
\begin{equation} \label{eq:averaged_input_signal_state}
\overline{\rho }=\int \mu (dm)\rho _{(\gamma ,m)}=
\rho _{(\gamma +Y_{\mu},0)}.
\end{equation}
That is, $\overline{\rho }$ is also a Gaussian state with the covariance matrix 
$\overline{\gamma }=\gamma +Y_{\mu }$ and has the vanishing displacement.
Equation~(\ref{eq:averaged_input_signal_state}) even holds for $Y_{\mu } \geq 0$.
Since the displacement of $\overline{\rho }$ is zero,
\begin{equation}
{\rm Tr}\overline{\rho }H =\frac{1}{2}\sum_{k=1}^{n}\omega _{k}\left(
\left\langle R_{2k-1}^{2}\right\rangle +\left\langle R_{2k}^{2}\right\rangle
\right)
=\frac{1}{4}\sum_{k=1}^{n}\omega _{k}{\rm Tr}\overline{\gamma }_{[k]},
\end{equation}
where $\overline{\gamma }_{[k]}$ denotes the principal submatrix of the $k$th mode of 
$\overline{\gamma }$ defined by Eq.~(\ref{eq:submatrix}) 
so that the energy constraint [(\ref{eq:energy_constraint_1})] is written as 
\begin{equation} \label{eq:energy_constraint_2}
\sum_{k=1}^{n}\omega _{k}{\rm Tr}(\gamma +Y_{\mu })_{[k]}=4\mathcal{E}.
\end{equation}
Since such a signal ensemble described above is not always optimal, we have
\begin{equation} \label{eq:Gaussian_Holevo_capacity_lower_bound_1}
C_{G}(\Phi ,\mathcal{E})\geq \sup_{\gamma, Y_{\mu }(\geq 0)}
\left[ S\left(\phi (\gamma +Y_{\mu
})\right)-S\left(\phi (\gamma )\right)\right], 
\end{equation}
where the supremum is taken under the constraint (\ref{eq:energy_constraint_2}) so that the right-hand side of Eq.~(\ref{eq:Gaussian_Holevo_capacity_lower_bound_1}) is written as
\begin{equation}
\sup_{\gamma }S(\phi [\gamma ])-\inf_{\gamma }S(\phi [\gamma ])
=\sup_{\gamma }S(\phi [\gamma ])-S_{\min }(\Phi ),
\end{equation}
where the supremum is taken under the constraint
\begin{equation} \label{eq:energy_constraint_3}
\sum_{k=1}^{n}{\rm Tr}\omega _{k}\gamma _{[k]}=4\mathcal{E}.
\end{equation}
Here, we note the following extremal property of Gaussian states.
For a given covariance matrix $\gamma $, 
the von Neumann entropy is maximized for the Gaussian state \cite{HSH99,WGC}.
Therefore, for the Gaussian signal state $\rho _{(\gamma ,m)}$ and 
the probability measure $\mu (d\gamma ,dm)$, 
the quantity within the brackets of the right-hand side of Eq.~(\ref{eq:Gaussian_Holevo_capacity}) 
cannot exceed the value of the right-hand side of 
(\ref{eq:Gaussian_Holevo_capacity_lower_bound_1}).
Therefore, the equality holds in the inequality (\ref{eq:Gaussian_Holevo_capacity_lower_bound_1});
\begin{equation}
C_{G}(\Phi ,\mathcal{E})=\sup_{\gamma}
S\left(\phi (\gamma)\right)-S_{\min }(\Phi ).
\end{equation}
Again, the supremum is taken under the constraint [Eq.~(\ref{eq:energy_constraint_3})].

In the above arguments, 
we have assumed implicitly that there exists a Gaussian state with the covariance matrix satisfying Eq.~(\ref{eq:energy_constraint_3}).
Otherwise, $C_{G}(\Phi ,\mathcal{E})$ should be zero 
because $\mathcal{E}$ in the energy constraint (\ref{eq:energy_constraint_1}) would be smaller 
than the sum of the energies of the zero-point oscillations over the modes.

\section{Additivity and multiplicativity problems of Gaussian channels} \label{additivity_problems}

For the tensor product of the Gaussian channels, $\Phi =\bigotimes_{j=1}^{m}\Phi _{j}$, 
it is evident from the definition that
\begin{equation} \label{eq:output_norm_inequality}
\xi _{p}(\Phi )\geq \prod_{j=1}^{m}\xi _{p}(\Phi _{j}).
\end{equation}
If the equality holds in the inequality (\ref{eq:output_norm_inequality}), 
we say that the maximal output $p$ norm is multiplicative for Gaussian channels $\Phi _{j}$.
To show the multiplicativity of the maximal output $p$ norm, it suffices to show
\begin{equation} \label{eq:multiplicativity}
\inf_{\gamma _{p}}F_{p}\left[\nu\left(\phi (\gamma _{p})\right)\right]=\prod_{j=1}^{m}\inf_{\gamma
_{p}}F_{p}\left[\nu\left(\phi _{j}(\gamma _{p})\right)\right].
\end{equation}
In Eq.~(\ref{eq:multiplicativity}), $\gamma _{p}$ in the left-hand side of the equation is 
the covariance matrix of a pure Gaussian state on the composite Hilbert space 
$\mathcal{H}_{1}\otimes \cdots \otimes \mathcal{H}_{m}$, while 
$\gamma _{p}$ in the right-hand side of the equation is 
the covariance matrix of a pure Gaussian state on the Hilbert space $\mathcal{H}_{j}$.

By noting Eq.~(\ref{eq:derivative}), 
it follows that if the maximal output $p$ norm is multiplicative, 
then the minimal output entropy is additive.

Let $\mathcal{E}$ 
be the value for the energy constraint [Eq.~(\ref{eq:energy_constraint_1})] for 
the Gaussian channel $\Phi _{j}$ and $\mathcal{E}=\sum_{j=1}^{m}\mathcal{E}_{j}$.
From the definition, the Gaussian Holevo capacity of the tensor product channel 
is greater than or equal to the supremum of the sum of the Gaussian Holevo capacity 
of individual channels;
\begin{equation} \label{eq:Gaussian_Holevo_capacity_lower_bound_2}
C_{G}(\Phi ,\mathcal{E})\geq 
\sup_{\{\mathcal{E}_{j}\},\sum_{j=1}^{m}\mathcal{E}_{j}=\mathcal{E}}
\sum_{j=1}^{m}C_{G}(\Phi _{j},\mathcal{E}_{j}).
\end{equation}
Here, the supremum is taken over all possible combinations of $\mathcal{E}_{j}$ 
under the constraint $\sum_{j=1}^{m}\mathcal{E}_{j}=\mathcal{E}$.
If the equality holds in the inequality (\ref{eq:Gaussian_Holevo_capacity_lower_bound_2}), 
we say that the energy-constrained Gaussian Holevo capacity is additive 
for Gaussian channels $\Phi _{j}$.
Now let $\rho $ be a Gaussian state on the composite Hilbert space 
$\mathcal{H}_{1}\otimes \cdots \otimes \mathcal{H}_{m}$
and define 
$\rho _{j}={\rm Tr}_{\mathcal{H}_{1}\otimes \cdots 
\otimes \mathcal{H}_{j-1}\otimes \mathcal{H}_{j+1}\otimes \cdots 
\otimes \mathcal{H}_{m}}\rho $.
By noting the subadditivity of the von Neumann entropy \cite{OP}, 
$S(\rho )\leq \sum_{j=1}^{m}S(\rho _{j})$, 
we have
\begin{equation}
S\left(\phi (\gamma )\right)\leq \sum_{j=1}^{m}S\left(\phi _{j}(\gamma _{j})\right),
\end{equation}
where $\gamma _{j}$ denotes the covariance matrix of the Gaussian state $\rho _{j}$.
Therefore, if the minimal output entropy $S_{\min }(\Phi )$ is additive 
for the channels $\Phi _{j}$, then
\begin{equation}
C_{G}(\Phi ,\mathcal{E})\leq 
\sup_{\{\mathcal{E}_{j}\},\sum_{j=1}^{m}\mathcal{E}_{j}=\mathcal{E}}
\sum_{j=1}^{m}C_{G}(\Phi _{j}, \mathcal{E}_{j}).
\end{equation}
This implies the additivity of the energy-constrained Gaussian Holevo capacity
\begin{equation}
C_{G}(\Phi ,\mathcal{E})=\sup_{\{\mathcal{E}_{j}\},\sum_{j=1}^{m}\mathcal{E}_{j}=\mathcal{E}}
\sum_{j=1}^{m}C_{G}(\Phi _{j},\mathcal{E}_{j}).
\end{equation}

Serafini {\it et al.} \cite{SEW} proved that 
the Gaussian maximal output $p$ norm of a tensor product of 
identical single mode Gaussian channels and that of single mode channels described by 
$X_{i}$ and $Y_{i}$ [(\ref{eq:channel_covariance})] 
such that $\det X_{i}$ are identical and $Y_{i}>0$ for all $i$, 
are multiplicative under Gaussian state inputs for $p > 1$.
Consequently, the Gaussian minimal output entropy and energy-constrained Gaussian Holevo capacity 
are additive for such tensor product channels.

\section{A majorization relation on symplectic eigenvalues} \label{majorization}

\smallskip

{\em Lemma 1. --}
Let $A$ be a $2n \times 2n$ real symmetric positive-definite matrix ($A=A^{T}>0$).
Then
\begin{equation} \label{eq:Lemma_eq_1}
\min_{SJ_{n}S^{T}=J_{k}}{\rm Tr}SAS^{T}=2\sum_{j=1}^{k}\nu _{j}^{\uparrow
}(A), \quad 1\leq k\leq n.
\end{equation}
The minimum in Eq.~(\ref{eq:Lemma_eq_1}) is taken over all $2k \times 2n$ real matrices $S$ 
satisfying $SJ_{n}S^{T}=J_{k}$.

{\em Proof.}
First of all, we note that a $2k \times 2n$ matrix $S$ satisfying $SJ_{n}S^{T}=J_{k}$ 
is the first $2k$ rows of a symplectic transformation, $S_{n}\in {\rm Sp}(2n,\mathbb{R})$, 
and that $A$ can be written in the Williamson standard form to obtain
\begin{equation}
\min_{SJ_{n}S^{T}=J_{k}}{\rm Tr}SAS^{T}=\min_{S_{n}J_{n}S_{n}^{T}=J_{n}}%
\sum_{j=1}^{2k}(S_{n}D_{A}S_{n}^{T})_{j,j}, 
\end{equation}
where we have used the fact that a product of symplectic transformations is 
a symplectic transformation and have defined 
$D_{A}=\mathrm{diag}
\left[\nu _{1}^{\uparrow }(A),\nu _{1}^{\uparrow }(A),\ldots
,\nu _{n}^{\uparrow }(A),\nu _{n}^{\uparrow }(A)\right]$.
Here, we write $S_{n}$ in the Euler decomposition form [Eq.~(\ref{eq:Euler})] to obtain
\begin{eqnarray} \label{eq:Lemma_eq_2}
\sum_{j=1}^{2k}(S_{n}D_{A}S_{n}^{T})_{j,j} 
&=&\sum_{j=1}^{k}\sum_{l,m=1}^{n}P_{2l-1,2m-1}^{(j)}z_{l}z_{m}Q_{2l-1,2m-1}
\nonumber \\
&&+\sum_{j=1}^{k}\sum_{l,m=1}^{n}P_{2l-1,2m}^{(j)}z_{l}z_{m}^{-1}Q_{2l-1,2m}
\nonumber \\
&&+\sum_{j=1}^{k}\sum_{l,m=1}^{n}P_{2l,2m-1}^{(j)}z_{l}^{-1}z_{m}Q_{2l,2m-1}
\nonumber \\
&&+\sum_{j=1}^{k}\sum_{l,m=1}^{n}P_{2l,2m}^{(j)}z_{l}^{-1}z_{m}^{-1}Q_{2l,2m},
\end{eqnarray}
where $z_{j}\geq 1$ ($j=1,2,\ldots ,n$),
\begin{equation}
P_{l,m}^{(j)}=T_{2j-1,l}^{(1)}T_{2j-1,m}^{(1)}+T_{2j,l}^{(1)}T_{2j,m}^{(1)},
\end{equation}
and
\begin{equation}
Q_{l,m}=\sum_{p=1}^{n}\nu _{p}^{\uparrow }(A)(
T_{l,2p-1}^{(2)}T_{m,2p-1}^{(2)}+T_{l,2p}^{(2)}T_{m,2p}^{(2)})
\end{equation}
with $T^{(1)},T^{(2)}\in K(n)$.
Using Eqs.~(\ref{eq:isomorphism_1}) and (\ref{eq:isomorphism_2}) for $T^{(1)}$ and $T^{(2)}$, 
the elements of $P^{(j)}$ and $Q$ are computed through 
the elements of $n \times n$ unitary matrices $U$ and $V$ as follows:
\begin{equation} \label{eq:P_1}
P_{2l-1,2m-1}^{(j)}=P_{2l,2m}^{(j)}=u_{j,l}^{R}u_{j,m}^{R}+u_{j,l}^{I}u_{j,m}^{I},
\end{equation}
\begin{equation} \label{eq:P_2}
P_{2l-1,2m}^{(j)}=-P_{2l,2m-1}^{(j)}=u_{j,l}^{R}u_{j,m}^{I}-u_{j,l}^{I}u_{j,m}^{R},
\end{equation}
\begin{equation} \label{eq:Q_1}
Q_{2l-1,2m-1}=Q_{2l,2m}=\sum_{p=1}^{n}\nu _{p}^{\uparrow }(A)\left(
v_{l,p}^{R}v_{m,p}^{R}+v_{l,p}^{I}v_{m,p}^{I}\right),
\end{equation}
and
\begin{equation} \label{eq:Q_2}
Q_{2l-1,2m}=-Q_{2l,2m-1}=\sum_{p=1}^{n}\nu _{p}^{\uparrow }(A)\left(
-v_{l,p}^{R}v_{m,p}^{I}+v_{l,p}^{I}v_{m,p}^{R}\right).
\end{equation}
Substituting Eqs.~(\ref{eq:P_1}), (\ref{eq:P_2}), (\ref{eq:Q_1}), and (\ref{eq:Q_2}) into Eq.~(\ref{eq:Lemma_eq_2}) yields 
\begin{eqnarray} \label{eq:Lemma_eq_3}
&&\sum_{j=1}^{2k}(S_{n}D_{A}S_{n}^{T})_{j,j}  \nonumber \\
&=&\frac{1}{4}\sum_{j=1}^{k}%
\sum_{l,m=1}^{n}u_{j,l}u_{j,m}^{*}(z_{l}+z_{l}^{-1})(z_{m}+z_{m}^{-1})%
\sum_{p=1}^{n}\nu _{p}^{\uparrow }(A)v_{l,p}v_{m,p}^{*}  \nonumber \\
&&+\frac{1}{4}\sum_{j=1}^{k}%
\sum_{l,m=1}^{n}u_{j,l}^{*}u_{j,m}(z_{l}+z_{l}^{-1})(z_{m}+z_{m}^{-1})%
\sum_{p=1}^{n}\nu _{p}^{\uparrow }(A)v_{l,p}^{*}v_{m,p}  \nonumber \\
&&+\frac{1}{4}\sum_{j=1}^{k}%
\sum_{l,m=1}^{n}u_{j,l}u_{j,m}^{*}(z_{l}-z_{l}^{-1})(z_{m}-z_{m}^{-1})%
\sum_{p=1}^{n}\nu _{p}^{\uparrow }(A)v_{l,p}^{*}v_{m,p}  \nonumber \\
&&+\frac{1}{4}\sum_{j=1}^{k}%
\sum_{l,m=1}^{n}u_{j,l}^{*}u_{j,m}(z_{l}-z_{l}^{-1})(z_{m}-z_{m}^{-1})%
\sum_{p=1}^{n}\nu _{p}^{\uparrow }(A)v_{l,p}v_{m,p}^{*}  \nonumber \\
&=&\sum_{l=1}^{4}\sum_{j=1}^{k}(W^{(l)}D_{A}W^{(l)\dagger })_{j,j},
\end{eqnarray}
where * means complex conjugate.
In the right-hand side of Eq.~(\ref{eq:Lemma_eq_3}), 
$W^{(1)}=UZ_{+}V$, 
$W^{(2)}=U^{*}Z_{+}V^{*}$, 
$W^{(3)}=UZ_{-}V^{*}$, 
and 
$W^{(4)}=U^{*}Z_{-}V$, 
with
\begin{equation}
Z_{\pm }=\frac{1}{2}\mathrm{diag}(z_{1}\pm z_{1}^{-1},\ldots ,z_{n}\pm z_{n}^{-1}).
\end{equation}
Note that matrices $W^{(l)}D_{A}W^{(l)\dagger }$ are positive semidefinite; 
$W^{(l)}D_{A}W^{(l)\dagger }\geq 0$ $(l=1,\ldots ,4)$.
Now let $\lambda _{j}^{\uparrow }(W^{(l)}D_{A}W^{(l)\dagger }) $ 
be eigenvalues of Hermitian matrices $W^{(l)}D_{A}W^{(l)\dagger }$.
By the Schur theorem (Appendix B), we have
\begin{equation}
\sum_{j=1}^{k}(W^{(l)}D_{A}W^{(l)\dagger })_{j,j}^{\uparrow }\geq
\sum_{j=1}^{k}\lambda _{j}^{\uparrow }(W^{(l)}D_{A}W^{(l)\dagger }),
\quad l=1,\ldots ,4.
\end{equation}
Thus, we obtain
\begin{eqnarray}
\sum_{j=1}^{2k}(S_{n}D_{A}S_{n}^{T})_{j,j} &\geq
&\sum_{l=1}^{4}\sum_{j=1}^{k}(W^{(l)}D_{A}W^{(l)\dagger })_{j,j}^{\uparrow }
\nonumber \\
&\geq &\sum_{l=1}^{4}\sum_{j=1}^{k}\lambda _{j}^{\uparrow
}(W^{(l)}D_{A}W^{(l)\dagger }).
\end{eqnarray}
Here,
\begin{equation}
\lambda _{j}^{\uparrow }(W^{(1)}D_{A}W^{(1)\dagger })=
\lambda _{j}^{\uparrow }(Z_{+}VD_{A}V^{\dagger }Z_{+})=
\lambda _{j}^{\uparrow }(Z_{+}CZ_{+})
\end{equation}
with $C=VD_{A}V^{\dagger }\geq 0$.
The first equality is due to the unitary invariance of the eigenvalues of the Hermitian matrices.
The eigenvalues, $\lambda _{j}^{\uparrow }(Z_{+}CZ_{+})$, 
admit the following max-min representation \cite{HJ}:
\begin{equation}
\lambda _{j}^{\uparrow }(Z_{+}CZ_{+})=
\max_{w_{1},\ldots ,w_{j-1}\in \mathbb{C}^{n}}
\min _{\substack{ x\neq 0  \\ x\perp w_{1},\ldots ,w_{j-1}}}
\frac{x^{\dagger }Z_{+}CZ_{+}x}{\left\| x\right\| _{2}^{2}}. 
\end{equation}
If we write $y=Z_{+}x$,
\begin{equation}
\left\| x\right\| _{2}^{2}=
\sum_{j=1}^{n}(Z_{+}^{-1})_{j}^{2}\left|y_{j}\right| ^{2}
\leq \sum_{j=1}^{n}\left| y_{j}\right| ^{2}=\left\|y\right\| _{2}^{2}. 
\end{equation}
Hence,
\begin{eqnarray}
\lambda _{j}^{\uparrow }(Z_{+}CZ_{+}) &\geq &
\max_{w_{1},\ldots ,w_{j-1}\in \mathbb{C}^{n}}
\min _{\substack{y\neq 0 \\ y\perp Z_{+}^{-1}w_{1},\ldots,Z_{+}^{-1}w_{j-1}}}
\frac{y^{\dagger }Cy}{\left\| y\right\| _{2}^{2}}
\nonumber \\
&=&\lambda _{j}^{\uparrow }(C)=\nu _{j}^{\uparrow }(A)
\end{eqnarray}
so that 
$\lambda _{j}^{\uparrow }(W^{(1)}D_{A}W^{(1)\dagger })\geq \nu _{j}^{\uparrow }(A)$.
Similarly, 
$\lambda _{j}^{\uparrow }(W^{(2)}D_{A}W^{(2)\dagger })\geq \nu _{j}^{\uparrow }(A)$.
Since 
$\lambda _{j}^{\uparrow }(W^{(3)}D_{A}W^{(3)\dagger })\geq 0$
and 
$\lambda _{j}^{\uparrow }(W^{(4)}D_{A}W^{(4)\dagger })\geq 0$,
we find
\begin{equation}
\sum_{j=1}^{2k}(S_{n}D_{A}S_{n}^{T})_{j,j}\geq 2\sum_{j=1}^{k}\nu
_{j}^{\uparrow }(A).
\end{equation}
The equality holds for $S_{n}=\mathbf{I}_{2n}\in {\rm Sp}(2n,\mathbb{R})$.
This completes the proof.
\hfill $\Box$

\smallskip

{\em Theorem 1. --}
Let $A$ and $B$ be $2n \times 2n$ real positive symmetric matrices 
($A=A^{T}>0$, $B=B^{T}>0$).
Then
\begin{equation} \label{eq:Theorem_eq_1}
\nu (A+B)\prec ^{w}\nu (A)+\nu (B).
\end{equation}

{\em Proof.}
By Lemma 1, we have
\begin{eqnarray} \label{eq:Theorem_eq_2}
2\sum_{j=1}^{k}\nu _{j}^{\uparrow }(A+B) &=&\min_{SJ_{n}S^{T}=J_{k}}
{\rm Tr}S(A+B)S^{T} \nonumber \\
&\geq &\min_{SJ_{n}S^{T}=J_{k}}{\rm Tr}SAS^{T}
+\min_{SJ_{n}S^{T}=J_{k}}{\rm Tr}SBS^{T} \nonumber \\
&=&2\sum_{j=1}^{k}\nu _{j}^{\uparrow }(A)+2\sum_{j=1}^{k}\nu _{j}^{\uparrow}(B) \nonumber \\
&=&2\sum_{j=1}^{k}\left[\nu (A)+\nu (B)\right]_{j}^{\uparrow }.
\end{eqnarray}
By definition of the weak supermajorization [(\ref{eq:weak_majorization})], 
inequality (\ref{eq:Theorem_eq_2}) yields the desired relation (\ref{eq:Theorem_eq_1}).
\hfill $\Box$

\section{Additivity and multiplicativity properties of Gaussian channels} \label{additivity_proof}

In this section, we focus on two classes of Gaussian channels; 
the classical noise channel and the thermal noise channel.
Both are important cases of Gaussian channels.

In the classical noise channel, 
a classical Gaussian noise is added to the input states.
Since 
$\mathcal{W}(\xi )R_{j}\mathcal{W}^{\dagger }(\xi )=R_{j}+\xi _{j}$,
the classical noise channel is described by
\begin{eqnarray}
\Phi ^{cl}(\rho _{(\gamma ,m)}) &=&\frac{1}{\pi ^{n}\sqrt{\det Y}}
\int d^{2n}\xi \exp (-\xi ^{T}Y^{-1}\xi )\mathcal{W}(\xi )\rho _{(\gamma ,m)}
\mathcal{W}^{\dagger }(\xi )  \nonumber \\
&=&\rho _{(\gamma +Y,m)}
\end{eqnarray}
with $Y\geq 0$.
Namely, the transformations of covariance matrix is given by
\begin{equation}
\phi ^{cl}(\gamma )=\gamma +Y.
\end{equation}

In the thermal noise channel, 
the signal Gaussian states interact with an environment that is in thermal equilibrium.
This channel is modeled by bemsplitters that couple the input Gaussian state 
and the thermal reservoir.
Let $a_{j}$ and $b_{j}$ be annihilation operators of the $j$th mode of the singnal state $\rho $ 
and the thermal state $\rho _{th}$ that acts as a thermal reservoir.
The action of the bemsplitter is described by the transformations, 
$a_{j}\mapsto \cos \theta _{j} a_{j}+\sin \theta _{j} b_{j} $ and 
$b_{j}\mapsto -\sin \theta _{J} a_{j}+\cos \theta _{j} b_{j} $.
Accordingly, the corresponding symplectic transformation takes the form,
\begin{equation} \label{eq:beam_splitter_of_thermal_noise}
S_{j}=\left( 
\begin{array}{cc}
\cos \theta _{j}\mathbf{I}_{2} & \sin \theta _{j}\mathbf{I}_{2} \\ 
-\sin \theta _{j}\mathbf{I}_{2} & \cos \theta _{j}\mathbf{I}_{2}
\end{array}
\right).
\end{equation}
Therefore, the output Gaussian state has the covariance matrix,
\begin{equation} \label{eq:thermal_noise_channel_covariance_1}
\phi ^{th}(\gamma )={\rm Tr}_{th}[S^{-1}(\gamma \oplus \gamma _{th})(S^{-1})^{T}],
\end{equation}
where 
$S=S_{1}\oplus S_{2}\oplus \cdots \oplus S_{n}$, 
$\gamma _{th}=\bigoplus_{j=1}^{n}( 2\left\langle n\right\rangle
_{j}+1) \mathbf{I}_{2}$ 
denotes the covariance matrix of the thermal state $\rho _{th}$ with 
$\left\langle n\right\rangle _{j}$ being the averaged photon number of the $j$th mode, 
and ${\rm Tr}_{th}$ describes the trace over the thermal state.
Using Eq.~(\ref{eq:beam_splitter_of_thermal_noise}), 
the right-hand side of Eq.~(\ref{eq:thermal_noise_channel_covariance_1}) is calculated as
\begin{equation} \label{eq:thermal_noise_channel_covariance_2}
\phi ^{th}(\gamma )=X^{T}\gamma X+Y
\end{equation}
where 
\begin{equation} \label{eq:thermal_noise_channel_covariance_X}
X=\bigoplus_{j=1}^{n}\sqrt{\eta _{j}}\mathbf{I}_{2}
\end{equation}
and
\begin{equation} \label{eq:thermal_noise_channel_covariance_Y}
Y=\bigoplus_{j=1}^{n}( 2\left\langle n\right\rangle _{j}+1) (1-\eta _{j})\mathbf{I}_{2}
\end{equation}
with
$\eta _{j}=\cos ^{2}\theta _{j}$ being the transmittivity of the bemsplitter.
At zero temperature ($\left\langle n\right\rangle _{j}=0$), 
the thermal noise channel is reduced to the lossy or attenuation channel \cite{HW,EW}.

\subsection{Classical noise channels}

For the $l_{j}$-mode classical noise channel $\Phi _{j}^{cl}$, 
the covariance matrix $\gamma $ is transformed according to 
$\phi _{j}^{cl}(\gamma )=\gamma +Y_{j}$, with $Y_{j}\geq 0$.
The tensor product of $\Phi _{j}^{cl}$ is also a classical noise channel, 
and the covariance matrix of the output is given by
\begin{equation}
\phi ^{cl}(\gamma )=\bigoplus_{j=1}^{m}\phi _{j}^{cl}(\gamma )=\gamma +Y, 
\end{equation}
where $\gamma $ is the covariance matrix of the input Gaussian state and 
$Y=\bigoplus_{j=1}^{m}Y_{j}$.
Since $Y_{j}$ is not always strictly positive definite, 
we add $\varepsilon \mathbf{I}_{2l_{j}}$ to $Y_{j}$ ($\varepsilon >0$); 
$Y_{j}(\varepsilon )=Y_{j}+\varepsilon \mathbf{I}_{2l_{j}}>0$ ($j=1,\ldots,m$) 
so that we can apply the Williamson theorem to $Y_{j}(\varepsilon )$.
Here, we write $Y(\varepsilon )=\bigoplus_{j=1}^{m}Y_{j}(\varepsilon )$.
By Williamson theorem, there exists $S_{j}\in {\rm Sp}(2l_{j},\mathbb{R})$ such that
\begin{eqnarray}
S_{j}Y(\varepsilon )_{j}S_{j}^{T} &=&
\mathrm{diag}\left(y_{1}^{(j)}(\varepsilon ),y_{1}^{(j)}(\varepsilon ),\ldots,
y_{l_{j}}^{(j)}(\varepsilon ),y_{l_{j}}^{(j)}(\varepsilon )\right) \nonumber \\
&=&D_{Y_{j}}(\varepsilon ).
\end{eqnarray}
Here, we write $S=\bigoplus_{j=1}^{m}S_{j}$ so that 
$SY(\varepsilon )S^{T}=\bigoplus_{j=1}^{m}D_{Y_{j}}(\varepsilon )
=D_{Y}(\varepsilon )$.
By Theorem 1, we have
\begin{equation} \label{eq:classical_noise_channel_majorization}
\nu \left[\gamma +Y(\varepsilon )\right]=
\nu \left[S\gamma S^{T}+D_{Y}(\varepsilon )\right]
\prec ^{w}\nu (\gamma )+\nu \left(D_{Y}(\varepsilon )\right). 
\end{equation}
Since $F_{p}$ is increasing and Schur-concave, 
(\ref{eq:classical_noise_channel_majorization}) yields
\begin{equation}
F_{p}\left[\nu (\gamma +Y(\varepsilon ))\right]\geq
F_{p}\left(\nu (\gamma )+\nu \left(D_{Y}(\varepsilon )\right)\right).
\end{equation}
Here we can take the limit $\varepsilon \rightarrow 0$ to obtain
\begin{equation} \label{eq:classical_noise_channel_multiplicative}
\inf_{\gamma _{p}}F_{p}\left[\nu (\phi ^{cl}(\gamma ))\right]\geq
\inf_{\gamma _{p}}F_{p}\left(\nu (\gamma _{p})+\nu (Y)\right),
\end{equation}
where 
$\nu (Y)=\bigoplus_{j=1}^{m}\nu (Y_{j})$ 
with
\begin{eqnarray}
\nu (Y_{j}) &=&\lim_{\varepsilon \rightarrow 0}(y_{1}^{(j)}(\varepsilon
),y_{2}^{(j)}(\varepsilon ),\ldots ,y_{l_{j}}^{(j)}(\varepsilon )) \nonumber \\
&=&(y_{1}^{(j)},y_{2}^{(j)},\ldots ,y_{l_{j}}^{(j)}).
\end{eqnarray}
The infimum in the right-hand side of (\ref{eq:classical_noise_channel_multiplicative}) 
is achieved for $\nu (\gamma _{p})=(1,1,\ldots ,1,1)$ and the equality holds if 
$S\gamma S^{T}$ takes the Williamson standard form.
Namely, for the covariance matrix $\gamma _{p}$ such that 
$S\gamma _{p}S^{T}=\mathrm{diag}(1,1,\ldots ,1,1)$,
\begin{eqnarray}
\inf_{\gamma _{p}}F_{p}\left[\nu \left(\phi ^{cl}(\gamma _{p})\right)\right]
&=&\prod_{j=1}^{m}\prod_{k=1}^{l_{j}}f_{p}(1+y_{k}^{(j)})  \nonumber \\
&=&\prod_{j=1}^{m}\inf_{\gamma _{p}}F_{p}\left[\nu \left(\phi _{j}^{cl}(\gamma _{p})\right)\right].
\end{eqnarray}
That is, the maximal output $p$ norm is multiplicative.
Consequently, the Gaussian minimal output entropy and the Gaussian Holevo capacity are additive.
Note that $S$ defined above is the direct sum of local symplectic transformations and 
$\mathrm{diag}(1,1,\ldots ,1,1)$ is the covariance matrix of the pure separable state 
so that the optimal $\gamma _{p}=S^{-1}\mathrm{diag}(1,1,\ldots ,1,1)(S^{T})^{-1}$ 
is a separable pure state.
This obvervation also indicates the multiplicativity of the maximal output $p$ norm.

\subsection{Thermal noise channels}

For the $l_{j}$-mode thermal noise channel $\Phi _{j}^{th}$, 
the covariance is transformed according to
\begin{equation}
\phi _{j}^{th}(\gamma )=\phi _{j}^{(0)}(\gamma )+Y_{j},
\end{equation}
where
\begin{equation}
Y_{j}=\bigoplus_{k=1}^{l_{j}}(2\left\langle n\right\rangle _{k}+1)(1-\eta
_{k})\mathbf{I}_{2}
\end{equation}
and
\begin{equation}
\phi _{j}^{(0)}(\gamma )=X^{T}\gamma X+Y,
\end{equation}
with
$X=\bigoplus_{k=1}^{l_{j}}\sqrt{\eta _{k}}\mathbf{I}_{2}$, 
$Y=\bigoplus_{k=1}^{l_{j}}(1-\eta _{k})\mathbf{I}_{2}$ and 
$0\leq \eta _{j}\leq 1$.
The tensor product of $\Phi _{j}^{th}$ is a Gaussian channel and 
the covariance matrix of the output state is given by
\begin{equation}
\phi ^{th}(\gamma )=\bigoplus_{j=1}^{m}\phi _{j}^{th}(\gamma )=\phi ^{(0)}(\gamma )+Y,
\end{equation}
where $\phi ^{(0)}(\gamma )=\bigoplus_{j=1}^{m}\phi _{j}^{(0)}(\gamma )$ and 
$Y=\bigoplus_{j=1}^{m}Y_{j}$.
Again, we add $\varepsilon \mathbf{I}_{2n}$ to $Y$ ($\varepsilon >0$); 
$Y(\varepsilon )=Y+\varepsilon \mathbf{I}_{2n}$ to ensure $Y(\varepsilon )>0$ 
($n=\sum_{j=1}^{m}l_{j}$).
Accordingly, we write 
$\phi ^{th}(\gamma ,\varepsilon )=\phi ^{(0)}(\gamma )+Y(\varepsilon )$.
By Theorem 1, we have
\begin{equation} \label{eq:thermal_noise_channel_majorization}
\nu \left[\phi ^{th}(\gamma ,\varepsilon )\right]\prec ^{w}
\nu \left(\phi ^{(0)}(\gamma )\right)+\nu \left(Y(\varepsilon )\right).
\end{equation}
Since $F_{p}$ is increasing and Schur-concave, (\ref{eq:thermal_noise_channel_majorization}) yields
\begin{equation}
F_{p}\{\nu \left[\phi ^{th}(\gamma ,\varepsilon )\right] \} \geq
F_{p}\left[\nu \left(\phi ^{(0)}(\gamma )\right)+\nu\left(Y(\varepsilon )\right)\right],
\end{equation}
Here we can take the limit $\varepsilon \rightarrow 0$ to obtain
\begin{equation} \label{eq:thermal_noise_channel_inequality}
\inf_{\gamma _{p}}F_{p}\left[\nu \left(\phi ^{th}(\gamma _{p})\right)\right]\geq
\inf_{\gamma _{p}}F_{p}\left[\nu \left(\phi ^{(0)}(\gamma _{p})\right)+\nu (Y)\right].
\end{equation}
Since the channel $\phi ^{(0)}(\gamma ) $ is completely positive, 
the Gaussian state with the covariance matrix $\phi ^{(0)}(\gamma _{p}) $ is a physical state 
so that $\nu _{j}\left(\phi ^{(0)}(\gamma _{p})\right)\geq 1$ ($j=1,2,\ldots ,n$).
For $\gamma =\mathrm{diag}(1,1,\ldots ,1,1)$, $\nu _{j}\left(\phi ^{(0)}(\gamma )\right)=1$ 
and the equality holds in (\ref{eq:thermal_noise_channel_inequality}).
Hence,
\begin{eqnarray}
\inf_{\gamma _{p}}F_{p}\left[\nu \left(\phi ^{th}(\gamma _{p})\right)\right]
&=&\prod_{j=1}^{m}\prod_{k=1}^{l_{j}}f_{p}\left[1+2(1-\eta _{k})\left\langle
n\right\rangle _{k}\right]  \nonumber \\
&=&\prod_{j=1}^{m}\inf_{\gamma _{p}}F_{p}\left[\nu \left(\phi _{j}^{th}(\gamma _{p})\right)\right].
\end{eqnarray}
That is, the maximal output $p$ norm is multiplicative.
Consequently, the Gaussian minimal output entropy and 
the energy-constrained Gaussian Holevo capacity are additive.

\section{Concluding remarks} \label{conclusion}

We proved the multiplicativity of maximal output $p$ norm of classical noise channels 
and that of thermal noise channels of arbitrary modes for all $p>1$ 
under the assumption that the input signal states were Gaussian states.
As a direct consequence, we also proved the additivity of the minimal output entropy 
and the energy-constrained Holevo capacity for those Gaussian channels under Gaussian inputs.
A majorization relation on symplectic eigenvalues was of importance in the proof.

At present, very little is known about the inequalities related to symplectic eigenvalues 
of real positive-definite matrices.
Efforts to unveil such unknown relations would assist 
in the analysis of entropic quantities of Gaussian states 
and would also shed light on the properties of Gaussian state entanglement \cite{WGK} 
and secure communication via Gaussian channels \cite{NBC}.

\subsection*{Acknowledgments}

The author would like to thank Masahito Hayashi, Osamu Hirota, Masaki Sohma, and Xiang-Bin Wang 
for useful comments and discussions.
He is grateful to Hiroshi Imai for support.

\appendix

\section{Concavity of $\ln f_{p}(x)$}

It is readily seen that $f_{p}(x)$ is concave for $1\leq p\leq 2$ and is convex for $p\geq 2$.
Therefore, $\ln f_{p}(x)$ is concave for $1\leq p\leq 2$.
In order to show the concavity of $\ln f_{p}(x)$ for $p\geq 2$, 
we examine the second derivative of $\ln f_{p}(x)$;
\begin{equation}
\frac{d^{2}}{dx^{2}}\ln f_{p}(x)=-\frac{p}{f_{p}^{2}(x)}g_{p}(x),
\end{equation}
where $g_{p}(x)=4p(x^{2}-1)^{p-2}+f_{p}(x)f_{p-2}(x) $.
For $p\geq 2$, we find that $g_{p}(x)\geq 0$ so that $d^{2}\ln f_{p}(x)/dx^{2}\leq 0$.
That is, $\ln f_{p}(x)$ is concave for $p\geq 2$.
Thus, $\ln f_{p}(x)$ is concave for all $p\geq 1$.

\section{Majorization and Schur convexity}

In this Appendix, we present definitions and basic facts on majorization 
and the Schur convexity (concavity) used in this paper \cite{MO}.

For vectors, $x=(x_{1},x_{2},\ldots ,x_{n})$ and 
$y=(y_{1},y_{2},\ldots ,y_{n})$ ($x_{j},y_{j}\in \mathbb{R}$), 
we write $x\leq y$ if $x_{j}\leq y_{j}$ ($j=1,2,\ldots ,n$).
Let 
$x^{\downarrow }=(x_{1}^{\downarrow },x_{2}^{\downarrow },\ldots ,x_{n}^{\downarrow })$
denote the decreasing rearrangement of $x$, 
where 
$x_{1}^{\downarrow }\geq x_{2}^{\downarrow }\geq \cdots \geq x_{n}^{\downarrow }$.
Similarly, let 
$x^{\uparrow }=(x_{1}^{\uparrow },x_{2}^{\uparrow },\ldots ,x_{n}^{\uparrow})$ 
denote the increasing rearrangement of $x$, 
where 
$x_{1}^{\uparrow }\leq x_{2}^{\uparrow }\leq \cdots \leq x_{n}^{\uparrow }$.

We say that $x$ is majorized by $y$ and write $x\prec y$ if
\begin{equation}
\sum_{j=1}^{k}x_{j}^{\downarrow }\leq \sum_{j=1}^{k}y_{j}^{\downarrow },
\quad k=1,2,\ldots ,n
\end{equation}
with the equality for $k=n$.

We say that $x$ is weakly submajorized (weakly supermajorized) by $y$ and write 
$x\prec _{w} (\prec ^{w})y$ if
\begin{equation} \label{eq:weak_majorization}
\sum_{j=1}^{k}x_{j}^{\downarrow (\uparrow )}\leq (\geq
)\sum_{j=1}^{k}y_{j}^{\downarrow (\uparrow )},\quad k=1,2,\ldots ,n.
\end{equation}

It is easy to see that 
$x\prec y$ if and only if $x\prec _{w}y$ and $x\prec ^{w}y$.

A real-valued function $f$ defined on $\mathbb{R}^{n}$ is said to be increasing if 
$x\leq y\Rightarrow f(x)\leq f(y)$, while $f$ is said to be decreasing if $-f$ is increasing.

A real-valued function $f$ defined on $\mathbb{R}^{n}$ is said to be Schur-convex if
$x\prec y\Rightarrow f(x)\leq f(y)$ while $f$ is said to be Schur-concave if $-f$ is Schur-convex.

A real-valued function $f$ defined on $\mathbb{R}^{n}$ satisfied 
$x \prec _{w} (\prec ^{w}) y\Rightarrow f(x)\leq (\geq) f(y)$ if and only if $f$ is increasing and Schur-convex (concave).

Let $g$ be a continuous and nonnegative function on $\mathbb{R}$.
Then, $f(x)=\prod_{j=1}^{n}g(x_{j})$ is Schur-convex (concave) if and only if 
$\log g$ is convex (concave).

An application of majorization theory to matrix analysis is the following Schur theorem \cite{Bha}.
Let $A\in M_{n}(\mathbb{C})$ be an Hermitian matrix.
Let $\mathrm{diag}(A)$ denote the vector whose elements are the diagonal entries of $A$ and 
$\lambda (A)$ the vector whose coordinates are eigenvalues of $A$.
Then, $\mathrm{diag}(A)\prec \lambda (A)$.

%%%%%%%%%%%%%%%%%%%%%%%%%%%%%%%%%%%%%%%%%%%%%%%%%%%%%%%%%%%%%%%%%%%%%%%%%%%%%%%
%%
%%%%%%%%%%%%%%%%%%%%%%%%%%%%%%%%%%%%%%%%%%%%%%%%%%%%%%%%%%%%%%%%%%%%%%%%%%%%%%%

%%%%%%%%%%%%%%%%%%%%%%%%%%%%%%%%%%%%%%%%%%%%%%%%%%%%%%%%%%%%%%%%%%%%%%%%%%%%%%%
%%
%%%%%%%%%%%%%%%%%%%%%%%%%%%%%%%%%%%%%%%%%%%%%%%%%%%%%%%%%%%%%%%%%%%%%%%%%%%%%%%
\end{document}